\documentclass[12pt,aps]{revtex4}

\usepackage[a4paper]{geometry}
\usepackage{amsmath}
\usepackage{amssymb}
\usepackage[dvips]{graphicx}
\usepackage{psfig}

\begin{document}

\noindent
\begin{center}
{\large \bf
Evidence against a glass transition in the 10-state 
short range Potts glass}\\
Claudio Brangian$^1$, Walter Kob$^2$, and Kurt Binder$^1$\\
$^1$ Institut f\"ur Physik, 
Johannes-Gutenberg-Universit\"at Mainz, Staudinger Weg 7, \\
D-55099 Mainz, Germany \\
$^2$ Laboratoire des Verres, Universit\'e Montpellier II, F-34095 
Montpellier, France 

\end{center}

\vspace*{7mm}
\par
\noindent
\begin{center}
\begin{minipage}[h]{122mm}
We present the results of Monte Carlo simulations of two different
10-state Potts glasses with random nearest neighbor interactions on a
simple cubic lattice. In the first model the interactions come from a $\pm
J$ distribution and in the second model from a Gaussian one, and in both
cases the first two moments of the distribution are chosen to be equal to
$J_0=-1$ and $\Delta J=1$. At low temperatures the spin autocorrelation
function for the $\pm J$ model relaxes in several steps whereas the one
for the Gaussian model shows only one.  In both systems the relaxation
time increases like an Arrhenius law. Unlike the infinite range model,
there are only very weak finite size effects and there is no evidence
that a dynamical or a static transition exists at a finite temperature.

\end{minipage}
\end{center}

\vspace*{5mm}
\par

PACS numbers: 64.70.Pf, 75.10.Nr, 75.50.Lk

\vspace*{5mm}
\par

\noindent
In recent years generalized spin-glass-type models such as the $p$-spin
model with $p\geq 3$ or the Potts glass with $p>4$, have found a large
attention~\cite{1,2,3,4,5,6,7,8,9,crisanti00,10,11,12} as prototype models for
the structural glass transition \cite{13,14,15,16,17,18}. In the case
of infinite range interactions, i.e. mean field, these models can be
solved exactly and it has been shown that they have a dynamical as well
as a static transition at a temperature $T_D$ and $T_0$, respectively
\cite{1,2,3,4,5,6,7,8,9,crisanti00,10,11,12}. At $T_D$ the relaxation times
diverge but no singularity of any kind occurs in the static properties,
whereas at $T_0$ a nonzero static glass order parameter appears
discontinuously. Close to $T_D$ the time and temperature dependence of
the spin autocorrelation function is described by the same type of mode
coupling equations \cite{3,7} that have been proposed by the idealized
version of mode coupling theory (MCT) \cite{14,15} for the {\it structural}
glass transition which suggests a fundamental connection between these
rather abstract spin models and real structural glasses.

Now it is well known that for {\it real} glasses the divergence of the
relaxation times predicted by MCT is rounded off since thermally
activated processes, which are not taken into account by this version of
the theory, become important~\cite{14,15}. This is in contrast to the mean
field case because there these processes are completely suppressed since
the barriers in the (free) energy landscape become infinitely high if
$T<T_D$. To what extend the {\it static} transition that exists in mean
field can be seen also in the real system, is a problem whose answer is
still controversial~\cite{9,16,17,18}.

In the present paper, we investigate whether the transitions present
in the case of a mean field 10-state Potts glass are also found if the
interactions are short ranged. The mean field model has recently been
studied in great detail by means of Monte Carlo methods and the results
are in good qualitative agreement with the main predictions of
the one-step replica symmetry breaking theory \cite{1,2,4,6}, although
very strong finite size effects occur, the nature of which are not fully
understood \cite{11}.

The Hamiltonian of the short range model is
\begin{equation} 
{\mathcal{H}}=-\sum_{\langle i,j\rangle} J_{ij} (p \; \delta_{\sigma_i \sigma_j}-1) \; .
\label{eq1}
\end{equation}
The Potts spins $\sigma_i$ on the lattice sites $i$ of the simple cubic
lattice take $p$ discrete values, $\sigma_i \in \{1,2,\ldots,p\}$,
and the index $j$ runs over the six nearest neighbors of site $i$. The
exchange constants $J_{ij}$ are taken either from a bimodal distribution
\begin{equation} 
P(J_{ij}) =x \delta (J_{ij}-J)+(1-x) \delta (J_{ij}+J)
\label{eq2}
\end{equation}
or a Gaussian distribution
\begin{equation}
P(J_{ij}) = [\sqrt{2\pi} (\Delta J)]^{-1} \exp 
\{-(J_{ij}-J_0)^2/2(\Delta J)^2\}.
\label{eq3}
\end{equation}
In both cases the first two moments are
chosen to be $J_0=[J_{ij}]_{av}=-1$ and $(\Delta
J)^2=[J_{ij}^2]_{av}-[J_{ij}]^2_{av}=+1$. (For $p=10$, a sufficiently
negative $J_0$ is necessary to avoid ferromagnetic order and we
have indeed found that neither the magnetization nor the magnetic
susceptibility show any sign of ferromagnetic ordering \cite{12}).) For
the distribution given by Eq.~(\ref{eq2}) this choice means $J=\sqrt{2}$
and $x=(1-1/\sqrt{2})/2\approx 0.146$.  We carry out Monte Carlo
runs with the standard heat-bath algorithm~\cite{19} making up to
$10^8$ Monte Carlo steps (MCS) per spin (for equilibration as well as
production), with lattices of linear dimensions $L=6$, $L=10$ and $L=16$,
and using periodic boundary conditions. The average over the quenched
disorder $[\cdots]_{av}$ is realized by averaging over $100$ independent
realizations of the system for $L=6$ and $10$ and over 50 realizations
for $L=16$. For $T=1.3$ and 1.4 and $L=10$ only 10 realizations were
used. In the following we will set the Boltzmann constant $k_B\equiv 1$
and measure temperature in units of $\Delta J$.

\begin{figure} [tbp]
\centerline{
\psfig{figure=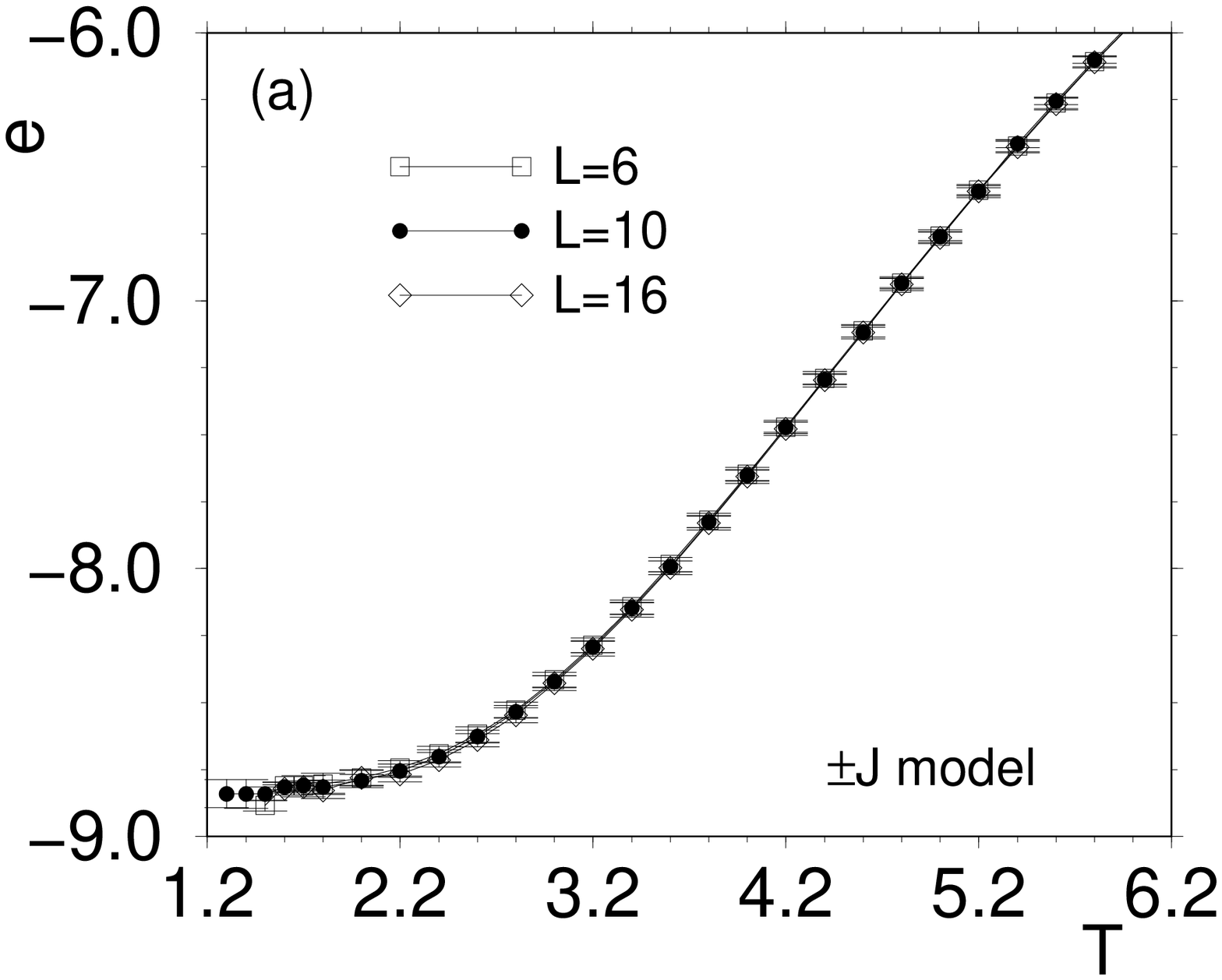,width=8.2cm,height=7.0cm}
\psfig{figure=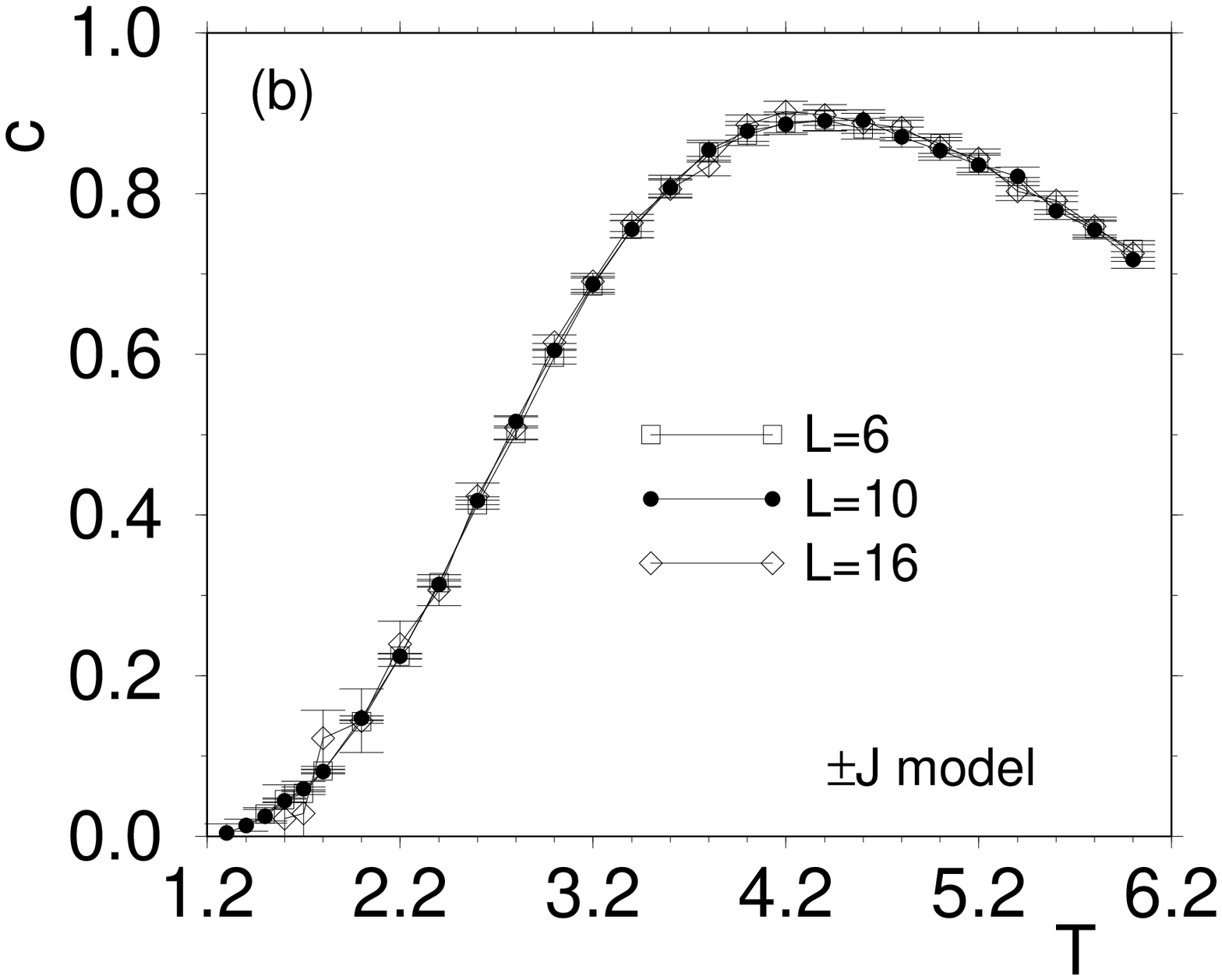,width=8.2cm,height=7.0cm}
}
\caption{Temperature dependence of the energy per spin $e(T)$, (a),
and specific heat per spin, (b). The different curves correspond to
different lattice sizes $L$.}

\label{fig1}
\end{figure}

In Fig.~\ref{fig1} we show the temperature dependence of the energy
per spin $e(T)$ and of the specific heat $c(T)$, for the model given
by Eq.~(\ref{eq2}). Both quantities seem to be essentially independent
of system size, in stark contrast to the results for the mean field
case~\cite{11}. Furthermore we see that for $T \leq 2$ the energy is
basically constant and we have found that at low $T$ the temperature
dependence of the specific heat is of the form $(\Delta/T)^2\exp(-\Delta
/T)$, with  $\Delta  \approx 14$, i.e. a dependence expected for a
two-level system with asymmetry $\Delta$. Thus we conclude that this data
does not show any evidence that in the temperature range investigated
a static transition occurs.

\begin{figure} [tbp]
\centerline{
\psfig{figure=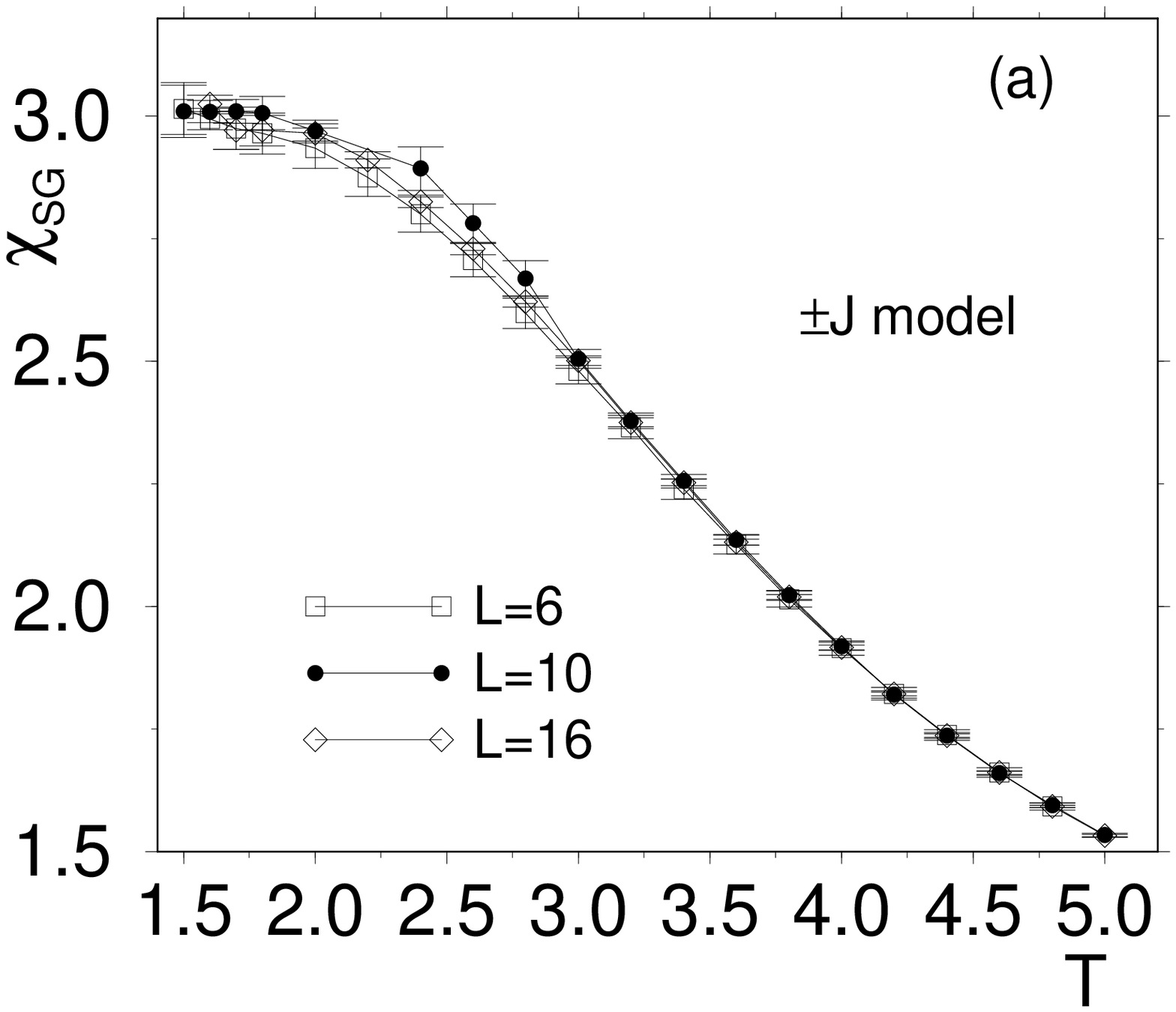,width=8.2cm,height=7.0cm}
\psfig{figure=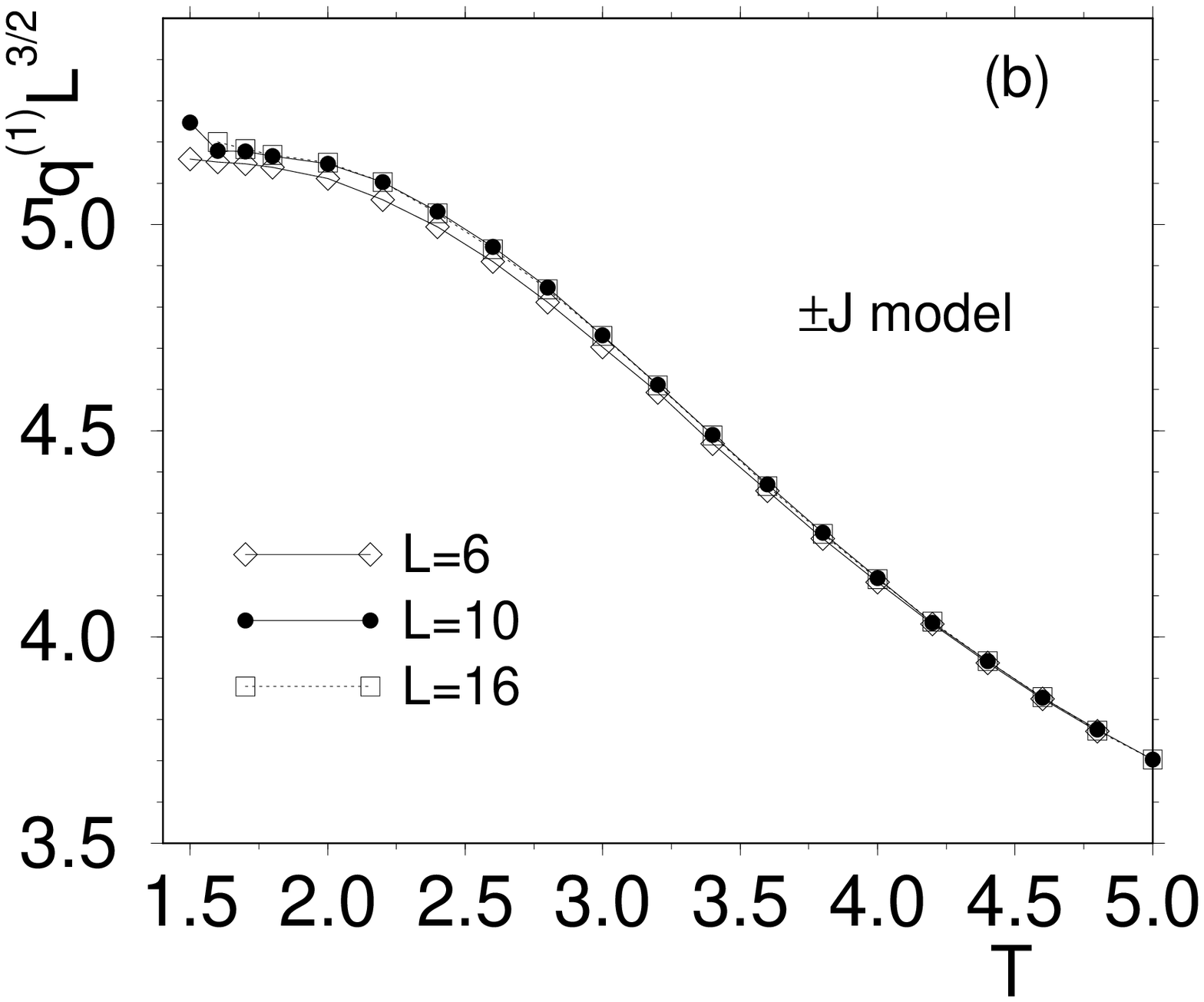,width=8.2cm,height=7.0cm}
}

\caption{Temperature dependence of the spin glass susceptibility
$\chi_{SG}$, (a), and scaled first moment of the order parameter
$[\langle q \rangle]_{av}$, (b). The different curves correspond
to different lattice sizes $L$.}

\label{fig2}
\end{figure}

A similar conclusion is reached from the $T-$dependence of the spin glass
susceptibility $\chi_{SG}=N[\langle q^2 \rangle]_{av}/(p-1)$. Here the
symmetrized order parameter $q$ is defined as \cite{10,11}
\begin{equation} 
q=\sqrt{\frac {1} {p-1} \, \sum\limits_{\mu, \nu=1}^{p-1} (q^{\mu \nu})^2 }  
\quad \mbox{with} \quad q^{\mu \nu} \equiv L^{-3}
\sum\limits_{i} (\vec{S}_{i, \alpha})^\mu ({\vec{S}_{i,\beta}})^\nu  ,
\quad \mu, \nu=1,2, \ldots , p-1 \quad ,
\label{eq4}
\end{equation}
and the spins $\vec{S}_{i,\alpha}$ refer to the simplex representation
of the Potts spin $\sigma_i$ in replica $\alpha$~\cite{simplex_ref}. (Note that
in order to define an order parameter it is necessary to consider two
replicas of the system, i.e. two systems with the same realization of the
disorder.)  Figure~\ref{fig2}a shows the $T-$ dependence of $\chi_{SG}$
and we see that this quantity is almost independent of $L$. Furthermore
we recognize that $\chi_{SG}$ remains finite even as $T \rightarrow
0$. Figure~\ref{fig2}b shows the (scaled) first moment of $q$ and we conclude
that it decreases like $1/\sqrt{L^3}$, i.e. shows a trivial size dependence
for all $T$. Also an analysis of fourth order cumulants (not shown here)
confirms the conclusion that no {\it static} transition occurs \cite{12}.
Similar results are found for the model with Gaussian distribution,
Eq.~(\ref{eq3}).

\begin{figure} [tbp]
\centerline{
\psfig{figure=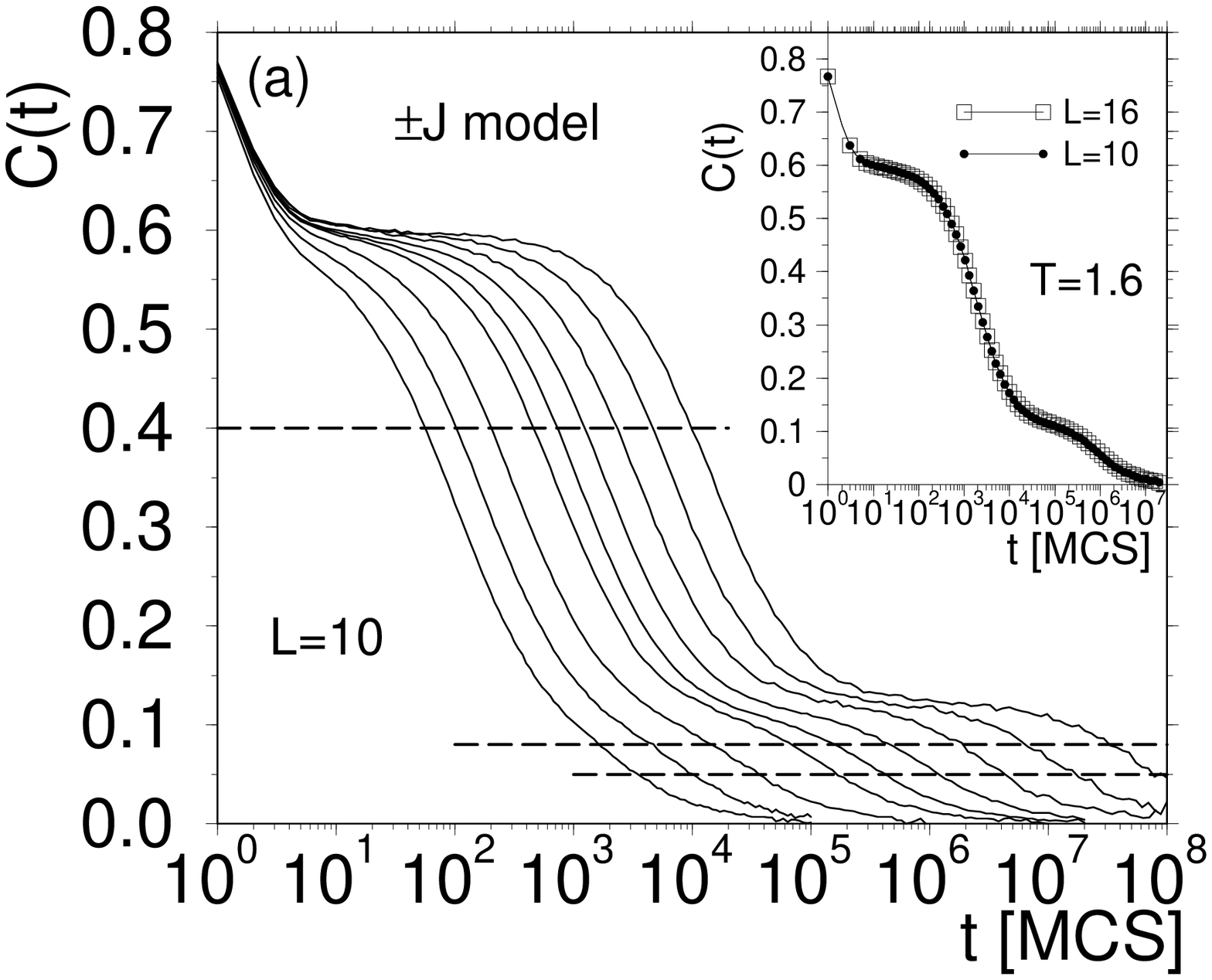,width=8.2cm,height=7.0cm}
\psfig{figure=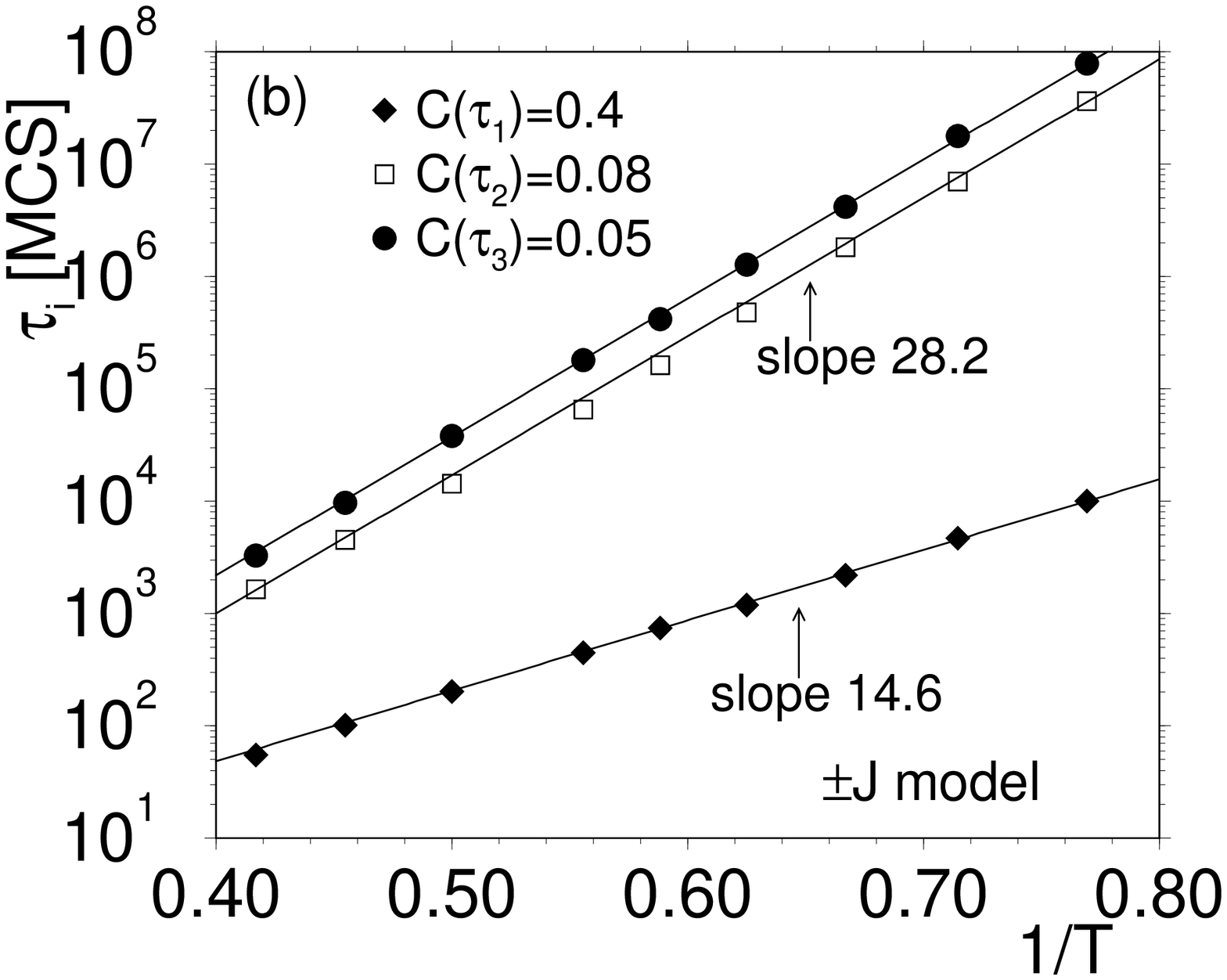,width=8.2cm,height=7.0cm}
}
\caption{(a) Time dependence of the spin autocorrelation function
$C(t)$ for temperatures (left to right) $T=2.4$, 2.2, 2.0, 1.8, 1.7,
1.6, 1.5, 1.4, and 1.3 ($L=10$). The horizontal dashed lines are used
to define the relaxation times $\tau_i$ (see text for details). In the
inset we compare $C(t)$ for $L=10$ and $L=16$ at $T=1.6$.  (b) Symbols:
Arrhenius plot of the relaxation time $\tau_i$ (see text for definition).
Straight lines: Fits to the data with an Arrhenius law. The numbers are
the activation energies.}

\label{fig3}
\end{figure}

In order to see whether this model has a {\it dynamic} transition at a
finite temperature we consider the time autocorrelation function of the
Potts spins:
\begin{equation}
C(t)=[L^3(p-1)]^{-1} \sum_{i=1}^{L^3}[\langle \vec{S}_i (t') \cdot \vec{S}_i (t' + t) 
\rangle]_{av} \quad .
\label{eq5}
\end{equation}
The time dependence of $C(t)$ is shown in Fig.~\ref{fig3}a for
various temperatures. We see that at intermediate temperatures
$C(t)$ has a plateau at around 0.6 with a width that increases
quickly with decreasing $T$. Such a behavior is very reminiscent
of the time and $T$ dependence of glass forming systems close to
the MCT temperature $T_D$~\cite{3,10,11,12,13,14,15,16}. However, as
we will show below, in this case the reason for the existence of a
plateau is very different. Interestingly enough $C(t)$ shows at low
temperatures a second plateau, and also its length increases rapidly
with decreasing temperature. Such a multi-step relaxation has so far
been seen only for few glass forming systems~\cite{garrahan00} and is
a rather unusual behavior. Below we will come back to this feature and
discuss its origin in more detail. That the time dependence of $C(t)$
is basically independent of the system size is demonstrated in the inset
of Fig.~\ref{fig3}a. Also this behavior differs strongly from the one
found for the mean field model~\cite{11}.

In order to study the slowing down of the dynamics of the system we can
define relaxation times $\tau_i$ via $C(\tau_1)=0.4$, $C(\tau_2)=0.08$
and $C(\tau_3)=0.05$. These definitions characterize the relaxation
times for the different processes seen in Fig.~\ref{fig3}a. The
temperature dependence of $\tau_i$ is shown in Fig.~\ref{fig3}b. We
see that this dependence can be described very well by an Arrhenius law
with an activation energy around 28.2 and 14.6 (straight lines). Such a
$T-$dependence is not in agreement with the expectation from MCT which
predicts around $T_D$ a power-law dependence. The observation that the
activation energies are close to $p|J|\approx 14.1$ and $2p|J|\approx
28.3$ suggests instead that the relaxation of the spins is given by
the breaking of one and two bonds, respectively. This interpretation is
also supported by the time dependence of the autocorrelation function
of {\it individual} spins~\cite{12}, since these functions typically
fall in three classes: Those that are relaxing fast (spins that have
only negative bonds), those that relax on intermediate times (one bond
needs to be broken) and those that relax slowly (two bonds that have to
be broken). Using these arguments and the concentration of ferromagnetic
bonds, $x$, one can estimate also the height of the plateaus~\cite{12}
which is predicted to be 0.61 and 0.13, which is in very good agreement
with the height that can be estimated from Fig.~\ref{fig3}a to be 0.59
and 0.12, respectively.

\begin{figure} [tpb]
\centerline{
\psfig{figure=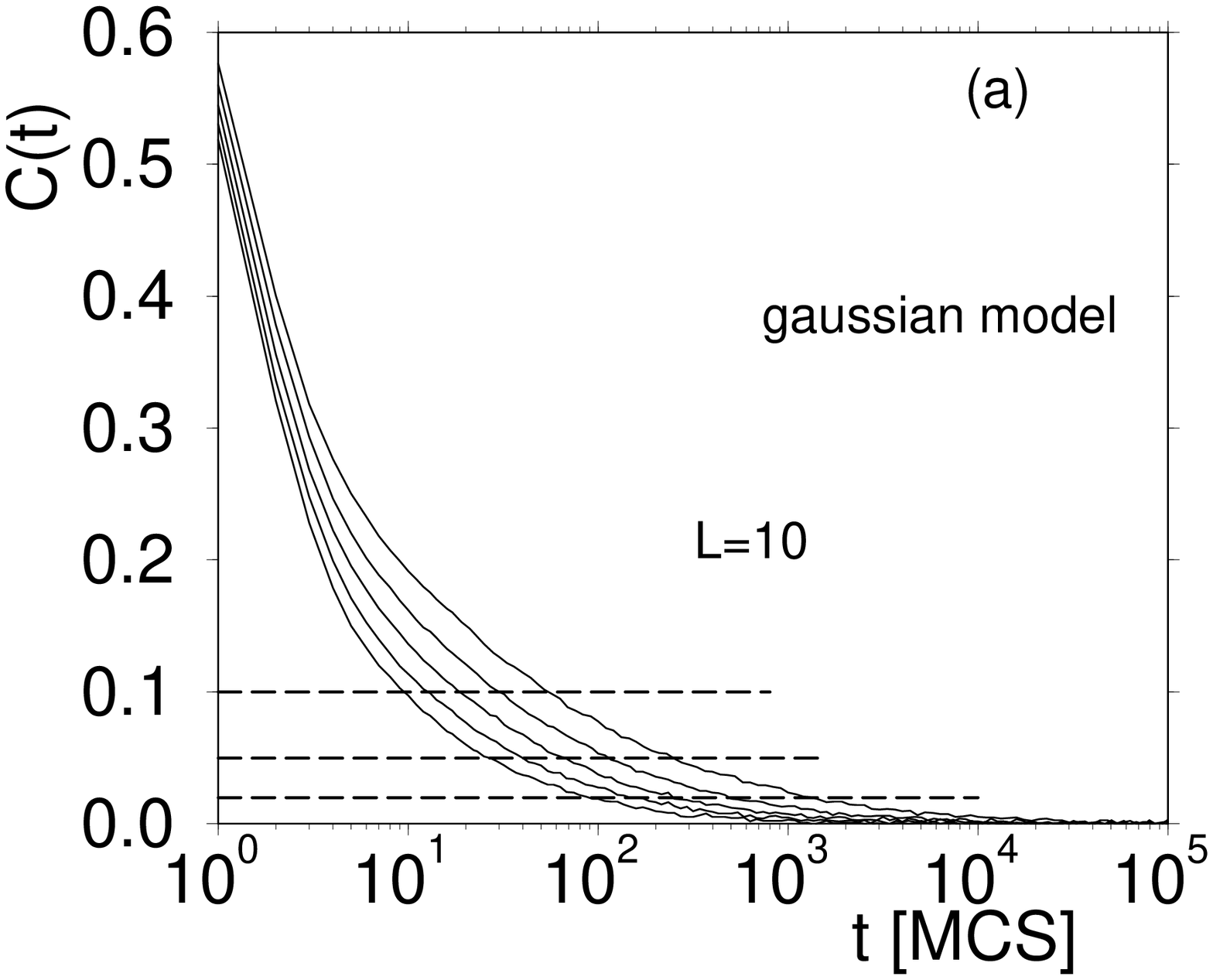,width=8.2cm,height=7.0cm}
\psfig{figure=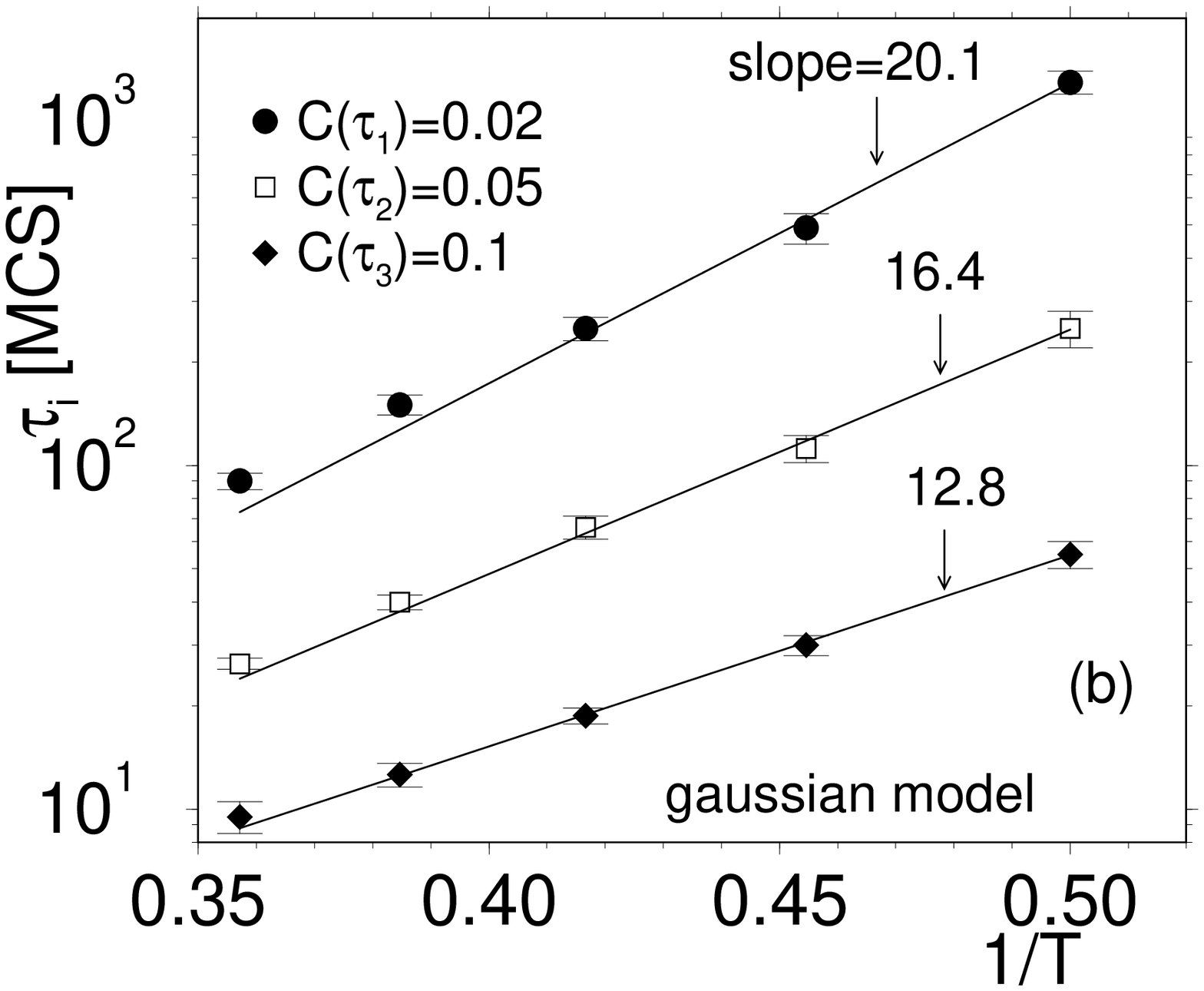,width=8.2cm,height=7.0cm}
}
\caption{(a) Time dependence of the spin autocorrelation function $C(t)$
for temperatures $T=2.8$, 2.6, 2.4, 2.2, and 2.0 (left to right) for
the Gaussian model. The horizontal dashed lines are used to define
the relaxation times $\tau_i$ (see text for details). (b) Symbols:
Arrhenius plot of the relaxation time $\tau_i$ (see text for definition).
Straight lines: Fits to the data with an Arrhenius law. The numbers are
the activation energies.}

\label{fig4}
\end{figure}

We emphasize that the occurrence of several plateaus in the
spin autocorrelation function (Fig.~\ref{fig3}a) is an immediate
consequence of the bond distribution, Eq.~(\ref{eq2}), and hence no
such plateaus are expected to occur for the Gaussian bond distribution,
Eq.~(\ref{eq3}). This expectation is indeed born out by the numerical
data shown in Fig.~\ref{fig4}a since $C(t)$ does not show any sign for
a plateau for any of the temperatures investigated. If one uses again
various definitions of relaxation times $\tau_i(T)$, this time the values
0.1, 0.05 and 0.02, their $T-$dependence is found again to be Arrhenius
like, see Fig.~\ref{fig4}b. However, in contrast to the results for the
$\pm J$ distribution the activation energies depend continuously on the
definition of $\tau_i$. Thus one finds that the dynamical transition
that the same model exhibits for the infinite range of interactions is
completely wiped out for the nearest neighbor case and we conclude that
neither a remnant of the static glass transition nor of the dynamic
transition exists.

Obviously the dynamical behavior of the present model is not similar
to the one of a supercooled liquid close to its glass transition. But
such a similarity can be expected to occur if we consider a variant
of the model that interpolates between the short range version of
the Potts glass and its infinite range version. E.g. we could choose
interactions that have a finite range but that extend further than
the nearest neighbors. For such a model we expect that a ``rounded''
version of the dynamic transition will occur, i.e., the autocorrelation
function develops a long-lived plateau and the relaxation time exhibits
the onset of a power law singularity, before it crosses over to a simple
Arrhenius behavior. This is the behavior found for large but {\it finite}
mean field systems~\cite{3,10,11,12}. Much less can be expected to be
seen in the static properties of that model: For any finite range of the
interaction, no nonzero order parameter can appear~\cite{stillinger88},
and a jump singularity in $\chi_{SG}$ cannot occur either. All what
remains is a rounded kink in the entropy vs.  temperature curve at
the rounded static transition, thus avoiding the Kauzmann \cite{20}
catastrophe of a vanishing configurational entropy. In fact,
this description is nicely consistent with all known facts about
the structural glass transition. Thus, if the analogy between the
latter and the behavior of medium range Potts glasses goes through,
the search for a {\it static} glass transition will remain elusive! This is
a somewhat surprising conclusion, since for $p=2$ the model reduces
to the Ising spin glass, where one knows that a (second-order) glass
transition temperature occurs at $T_c>0$ \cite{21}, and also for the
$p=3$ Potts glass there seems to be at least a divergent $\chi_{SG}$
as $T \rightarrow 0$ \cite{22,23}. However, for large values of $p$
the static and dynamic properties seem to be very different.

\bigskip

Acknowledgements: This work was supported by the Deutsche
Forschungsgemeinschaft (DFG) under grant N$^{\circ}$ SFB 262/D1. We
thank the John von Neumann Institute for Computing (NIC) at J\"ulich
for a generous grant of computing time at the CRAY-T3E.

\clearpage

\newpage


\begin{thebibliography}{22}

\bibitem{1}
ELDERFIELD, D. and SHERRINGTON, D.,
{\it J. Phys. C}, {\bf 16} (1983) L497, L971, L1169.

\bibitem{2}
GROSS, D. J., KANTER, I., and SOMPOLINSKY, H.  
{\it Phys. Rev. Lett.} {\bf 55} (1985) 304.

\bibitem{3}
KIRKPATRICK, T. R. and WOLYNES, P. G.,
{\it Phys. Rev. B} {\bf 36} (1987) 8552; 
%
KIRKPATRICK, T.R. and THIRUMALAI, D.,
{\it Phys. Rev. B} {\bf 37} (1988) 5342; 
%
THIRUMALAI, D. and KIRKPATRICK, T. R.,
{\it Phys. Rev. B} {\bf 38} (1988) 4881.

\bibitem{4}
CWILICH, G. and KIRKPATRICK, T. R.,
{\it J. Phys. A: Math.  Gen.} {\bf 22} (1989) 4971; 
%
CWILICH, G.,
{\it J. Phys. A: Math. Gen.} {\bf 23} (1990) 5029.

\bibitem{5}
CRISANTI, A., HORNER, H., and SOMMERS, H.-J.,
{\it Z. Phys. B} {\bf 92} (1993) 257.

\bibitem{6}
DE SANTIS, E., PARISI, G., and RITORT, F.,
{\it J. Phys. A: Math. Gen.} {\bf 28} (1995) 3025.

\bibitem{7}
KIRKPATRICK, T. R., and THIRUMALAI, D.,
{\it Transp. Theory Stat. Phys.} {\bf 24} (1995) 927.

\bibitem{8}
DILLMANN, O., JANKE, W. and BINDER, K.,
{\it J. Stat. Phys.} {\bf 92} (1998) 57.

\bibitem{9}
FRANZ, S. and PARISI, G.,
{\it Physica A} {\bf 261} (1998) 317; 
%
MEZARD, M. and PARISI, G.,
{\it Phys. Rev. Lett.} {\bf 82} (1999) 317; 
%
PARISI, G.,
{\it Physica A} {\bf 280} (2000) 115.

\bibitem{crisanti00}
CRISANTI A. and RITORT F.,
{\it Physica A} {\bf 280} (2000) 155.

\bibitem{10}
BRANGIAN, C., KOB, W., and BINDER, K.,
{\it Europhys. Lett.} {\bf 53} (2001) 756.

\bibitem{11}
BRANGIAN, C., KOB, W., and BINDER, K.,
{\it J. Phys. A: Math.  Gen.} {\bf 35} (2002) 191.

\bibitem{12}
BRANGIAN, C. 
{\it Dissertation (Johannes Gutenberg Universit\"at, Mainz)} (2002).

\bibitem{13}
J\"ACKLE, J.,
{\it Rep. Progr. Phys.} {\bf 49} (1986) 171.

\bibitem{14}
G\"OTZE, W. 
in {\it Liquids, Freezing, and Glass Transition} Eds.: Hansen, J. P., Levesque, D., and
Zinn-Justin, J. (North-Holland, Amsterdam) 1990 p. 287.

\bibitem{15}
G\"OTZE, W.,
{\it J. Phys.: Condens. Matter} {\bf 11} (1999) A1.

\bibitem{16}
KOB, W.,
{\it J. Phys.: Condens. Matter} {\bf 11} (1999) R85.

\bibitem{17}
BINDER, K.,
{\it J. Non-Cryst. Solids} {\bf 274} (2000) 332.

\bibitem{18}
BRANGIAN, C., KOB, W., and BINDER, K.,
{\it Proceedings of the NATO ARW on ``New Kinds of Phase
Transitions in Disordered Materials''} (in press).

\bibitem{19}
LANDAU, D. P. and BINDER, K.,
{\it A Guide to Monte Carlo Simulation in Statistical Physics}
(Cambridge Univ. Press, Cambridge) 2000.

\bibitem{simplex_ref}
WU F. Y.,
{\it Rev. Mod. Phys.} {\bf 54} (1982) 235;
%
{\it Rev. Mod. Phys.} {\bf 55} (1982) 315;
%
ZIA R. K. and WALLACE D. J.,
{\it J. Phys. A: Math Gen.} {\bf 8} (1975) 1495.

\bibitem{garrahan00}
See, e.g., GARRAHAN J. P. and NEWMAN M. E. J., 
{\it Phys. Rev. E} {\bf 62} (2000) 7670.

\bibitem{stillinger88}
STILLINGER F. H., 
{\it J. Chem. Phys.} {\bf 88}, (1988) 7818.

\bibitem{20}
KAUZMANN, W.,
{\it Chem. Rev.} {\bf 43} (1948) 219.

\bibitem{21}
BINDER, K. and YOUNG, A. P.,
{\it Rev. Mod. Phys.} {\bf 58} (1986); 
%
YOUNG, A. P.,
{\it Spin Glasses and Random Fields} (World Scientific, Singapore)  1998.

\bibitem{22}
SCHEUCHER, M. and REGER, J. D.,
{\it Z. Phys. B} {\bf 91} (1993) 383.

\bibitem{23}
BINDER, K. and REGER, J. D.,
{\it Adv. Phys.} {\bf 41} (1992) 547.

\end{thebibliography}
\end{document}